\documentclass[twocolumn,pre,superscriptaddress,aps,footinbib]{revtex4}
\usepackage{url,psfrag,graphicx}
\usepackage{dcolumn}
\usepackage{amsmath,amssymb}
\usepackage{bm}
\usepackage{pstricks,multido}
\usepackage{hyperref}
\usepackage{float,epsfig,color}
\usepackage{collref}
\usepackage{times}
\def\pl{\partial}

\def\<{\left<}
\def\>{\right>}
\def\ket|#1>{\left|#1\right>}
\def\bra<#1|{\left<#1\right|}
\def\elem<#1|#2|#3>{\left<#1\right|#2\left|#3\right>}
\def\({\left(}
\def\){\right)}

\psset{unit=1mm}
\def\beq{\begin{equation}}
\def\eeq{\end{equation}}

\begin{document}

\author{Silvia N.\ Santalla}
\affiliation{Departamento de F\'{\i}sica and Grupo Interdisciplinar de Sistemas
  Complejos (GISC), Universidad Carlos III de Madrid, Legan\'es, Spain}

\author{Javier Rodr\'{\i}guez-Laguna}
\affiliation{Departamento de F\'{\i}sica Fundamental, Universidad Nacional de
  Educaci\'on a Distancia (UNED), Madrid, Spain}

\author{Jos\'e P.\ Abad}
\affiliation{Departamento de Biolog\'{\i}a Molecular, Universidad
  Aut\'onoma de Madrid (UAM), Cantoblanco, Madrid, Spain}

\author{Irma Mar\'{\i}n}
\affiliation{Departamento de Biolog\'{\i}a Molecular, Universidad
  Aut\'onoma de Madrid (UAM), Cantoblanco, Madrid, Spain}

\author{Mar\'{\i}a del Mar Espinosa}
\affiliation{Hospital Universitario Puerta de Hierro, Madrid, Spain}

\author{Javier Mu\~noz-Garc\'{\i}a}
\affiliation{Departamento de Matem\'aticas \& GISC, Universidad Carlos III de
  Madrid, Legan\'es, Spain}

\author{Luis V\'azquez}
\affiliation{Instituto de Ciencia de Materiales de Madrid (ICMM), Consejo Superior de Investigaciones Científicas (CSIC),
  Madrid, Spain}

\author{Rodolfo Cuerno}
\affiliation{Departamento de Matem\'aticas \& GISC, Universidad Carlos III de
  Madrid, Legan\'es, Spain}

\title[Short Title]{Nonuniversality of front fluctuations for compact
  colonies of nonmotile bacteria}


\begin{abstract}
The front of a compact bacterial colony growing on a Petri dish is a
paradigmatic instance of non-equilibrium fluctuations in the
celebrated Eden, or Kardar-Parisi-Zhang (KPZ), universality
class. While in many experiments the scaling exponents crucially
differ from the expected KPZ values, the source of this disagreement
has remained poorly understood. We have performed growth experiments
with {\em B.\ subtilis} 168 and {\em E.\ coli} ATCC 25922 under
conditions leading to compact colonies in the classically-alleged Eden
regime, where individual motility is suppressed. Non-KPZ scaling is
indeed observed for all accessible times, KPZ asymptotics being ruled
out for our experiments due to the monotonic increase of front
branching with time. Simulations of an effective model suggest the
occurrence of transient non-universal scaling due to diffusive
morphological instabilities, agreeing with expectations from detailed
models of the relevant biological reaction-diffusion processes.
\end{abstract}


\maketitle


\section{Introduction}

Active matter, i.e., the emergent behavior of a large number of agents
that can produce mechanical forces via energy dissipation
\cite{Ramaswamy_10}, is recently proving itself as an extremely rich
context for non-equilibrium phenomena. Instances range from schools of
fish or bird flocks, to vibrated granular rods or propelled
nanoscale or colloidal particles, for all of which fluctuations play a
conspicuous role in the collective dynamics \cite{Marchetti_13}.

Bacterial systems \cite{Ben-Jacob_00} provide further instances of
active matter, from microswimmer suspensions in which single cell
motility plays a crucial role \cite{Sokolov_07,Zhang_10} to bacterial
colonies, in which motility can be hampered
\cite{Ben-Jacob_98,Matsushita_04,Bonachela_11}. Actually, the fronts
of bacterial colonies have long been held as textbook examples
\cite{Vicsek_92,Barabasi_95,Meakin_98} on how interactions among
individuals lead to collective, highly-correlated behavior. For
experiments frequently done using {\em Bacillus subtilis} or {\em
  Escherichia coli}, this ranges from the formation of characteristic
patterns ---like diffusion-limited aggregation (DLA) fractals,
concentric rings, or dense-branched morphologies--- to formation of
disks or of compact, but rough, morphologies
\cite{Fujikawa_89,Vicsek_90,Wakita_97,Matsushita_98,Matsushita_04},
all of which are also found in other, non-living, systems.

The simplest situation in which individual bacterial motility is fully
suppressed by a high agar concentration on the supporting Petri dish
has received particular attention, as it paradigmatically demonstrates
a change from DLA branches to compact, Eden-like, clusters, for an
increasing nutrient concentration \cite{Fujikawa_89,Bonachela_11},
akin to that found for many other diffusion-limited (DL) growth
systems \cite{Nicoli_09}. This morphological transition has been
recently shown to bear direct importance on the biological performance
of the colony \cite{Nadell_10,Mitri_11,Nadell_13}: branches
enable the space segregation of cell lines which respond differently
with respect to the production of enzymes needed for biofilm
formation, enhancing the prevalence of cooperative cells. Biofilms are
surface-attached communities hosting most living bacteria in nature,
of paramount importance to medicine and technology, from infections to
energy harvesting \cite{Costerton_95,Wilking_11}.

Furthermore, front fluctuations of Eden clusters \cite{Eden_61} are in
the celebrated Kardar-Parisi-Zhang (KPZ) \cite{Kardar_86} universality
class of kinetic roughening
\cite{Barabasi_95,Alves_11,Takeuchi_12}. Sparked by breakthroughs on
exact solutions of the KPZ equation and related growth models, that
have been experimentally validated (see \cite{Halpin-Healy_15} for a review),
this class is recently being demonstrated as a paradigm for strong fluctuations in one dimension (1D), as found e.g.\ in non-linear oscillators \cite{VanBeijeren_12}, stochastic hydrodynamics \cite{Mendl_13}, quantum liquids \cite{Kulkarni_13}, 
or reaction-diffusion systems \cite{Nesic_14}. Remarkably,
in the low motility case, most experimental values
found for the scaling exponents of compact Eden-like
bacterial colonies {\em do not} coincide with the KPZ values
\cite{Vicsek_90,Wakita_97,Bonachela_11}. This fact has been reconciled with a
putative Eden behavior via e.g.\ effective quenched
disorder \cite{Bonachela_11}, unexpectedly for a system which is
succesfully described by continuum
\cite{Lacasta_99,Mimura_00,Dockery_02,Kobayashi_04,Giverso_15,Giverso_16}
or discrete \cite{Nadell_10,Farrell_13,Farrell_17} models with no
source of quenched disorder.

In this article, we report colony growth experiments for {\em
B.\ subtilis} and {\em E.\ coli} under suppressed-motility conditions  \cite{Matsushita_98,Rafols_98} in the alleged Eden regime. We explain the
non-KPZ kinetic roughening that we indeed observe as non-universal
scaling behavior induced by the diffusive instabilities that
occur. This is achieved by comparing our data with simulations of a
continuum model that we formulate, indicating that these experimental
conditions keep the system within a DL transient for all accessible
times. Moreover, the increase of front branching with time for the experimental colonies prevents asymptotics from being in the KPZ universality class under our suppressed-motility conditions. Analogous non-universal behavior has been identified in other DL systems, like thin film growth by electrodeposition (ECD), by chemical vapor deposition (CVD)
\cite{Castro_00,Nicoli_09}, or in coffee ring formation by evaporating
colloidal suspensions \cite{Yunker_13,Nicoli_13,Yunker_13_2,Oliveira_14}.

The paper is organized as follows. Our experimental setup and methods are described in Sec.\
\ref{sec:experiments}, while a continuum model which we employ to rationalize our observations is described in Sec.\ \ref{sec:model}. This is followed by our experimental results, which are reported in Sec.\ \ref{sec:analysis}. Further discussion is provided in Sec.\ \ref{sec:conclusions}, which also contains our conclusions and an outlook on future work. Further technical details on error estimates are left to an appendix.


\section{Experimental Setup}
\label{sec:experiments}

We have grown colonies of {\em B.\ subtilis} 168 (BS) and {\em
  E.\ coli} ATCC 25922 (EC) on Petri dishes as in
\cite{Matsushita_98,Rafols_98}, in the high agar concentration (i.e.,
low motility) regime for different concentrations of
nutrients. Specifically, we have kept a constant agar concentration
$C_a=10$ g/l while considering different values of the nutrient
concentration, $C_n = 10, 15$, or $20$ g/l, within the Eden-like
region in the morphological space of
\cite{Matsushita_98,Rafols_98}. These conditions correspond to a value
for the non-dimensional thickness $\delta$ of the active layer within
the bacterial colony, where the nutrient concentration has
non-negligible gradients
\cite{Dockery_02,Nadell_10,Farrell_13,Giverso_16,Farrell_17}, which is
large enough for the colony to look compact on the accesible
space-time scales.

For inoculating Petri dishes, bacteria were grown overnight in
nutritive liquid medium [5 g/l NaCl (Merck, Germany), 3 g/l meat
extract (Merck, Germany), 10 g/l bacto-peptone (Lab.\ Conda, Spain)]
and the OD600 was measured. Cells were pelleted at 12 krpm in a
microcentrifuge, and resuspended to 0.5 OD600 in minimal medium
without bacto-peptone. Two replica Petri dishes were prepared
following \cite{Rafols_98}: a 3 mm thick agar plate in nutritive
medium [5 g/l NaCl (Merck, Germany), 5 g/l K${}_2$HPO${}_4$ (Carlo
Erba, Italy) and bacto-peptone (Lab.\ Conda, Spain)] inoculated at
the center with 1 $\mu$l of the cell suspension was incubated at 35
${}^\circ$C in a sealed humid chamber for up to 33 days, leading to
growth of quasi-2D colonies. No swarming of bacteria has been
detected.

\begin{figure}
\centering
\mbox{
  \includegraphics[width=0.25\textwidth]{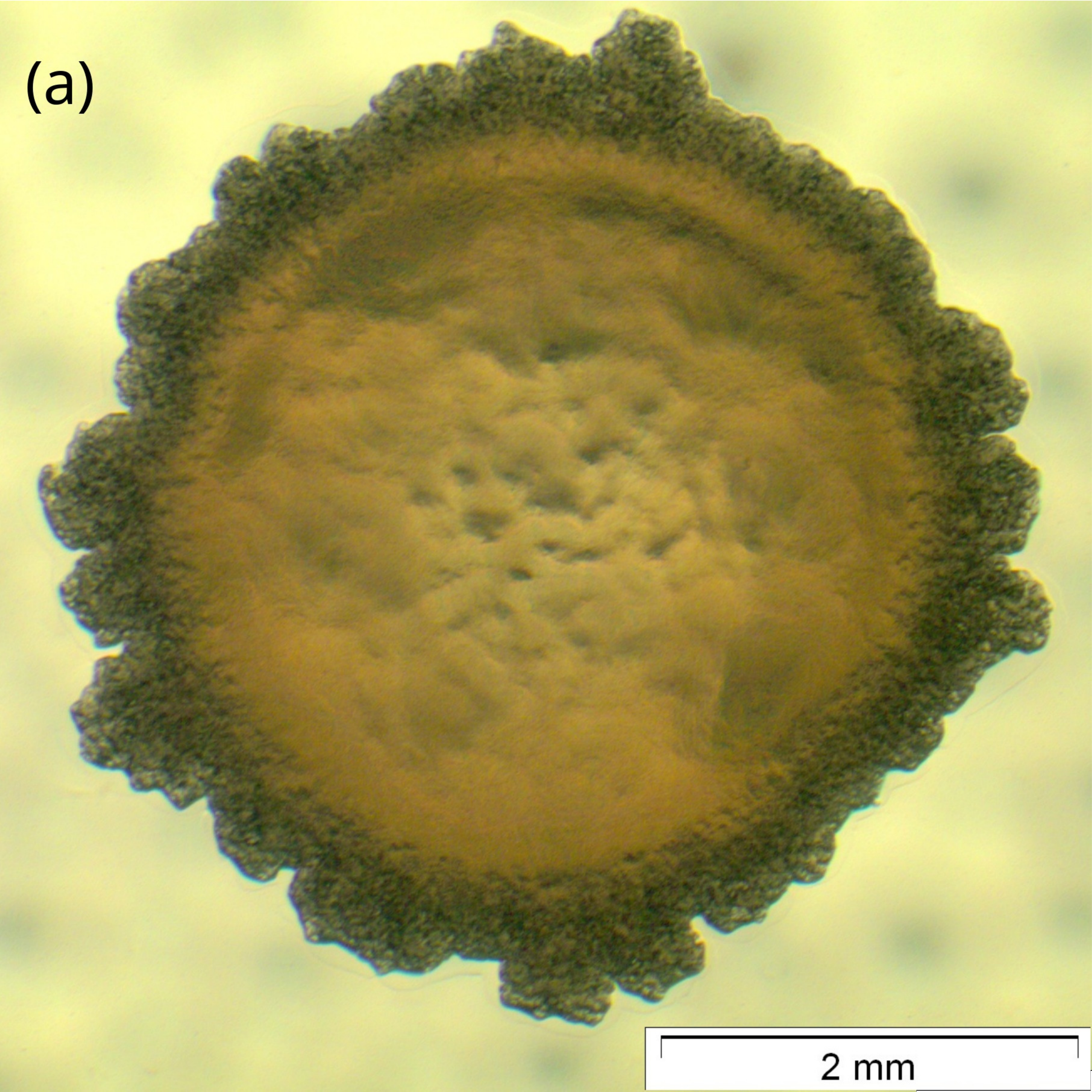}
  \includegraphics[width=0.25\textwidth]{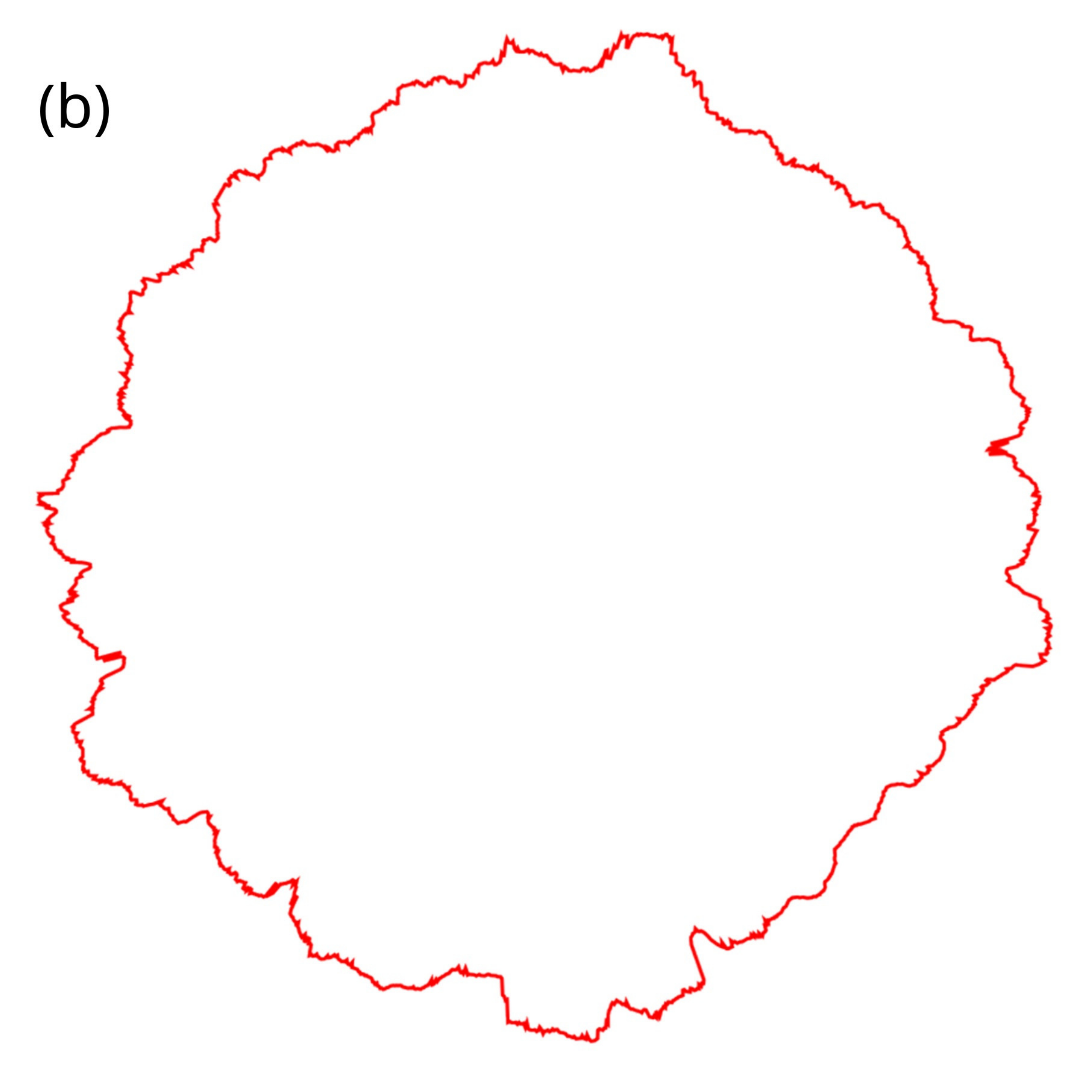}}
\\
\mbox{
  \includegraphics[width=0.25\textwidth]{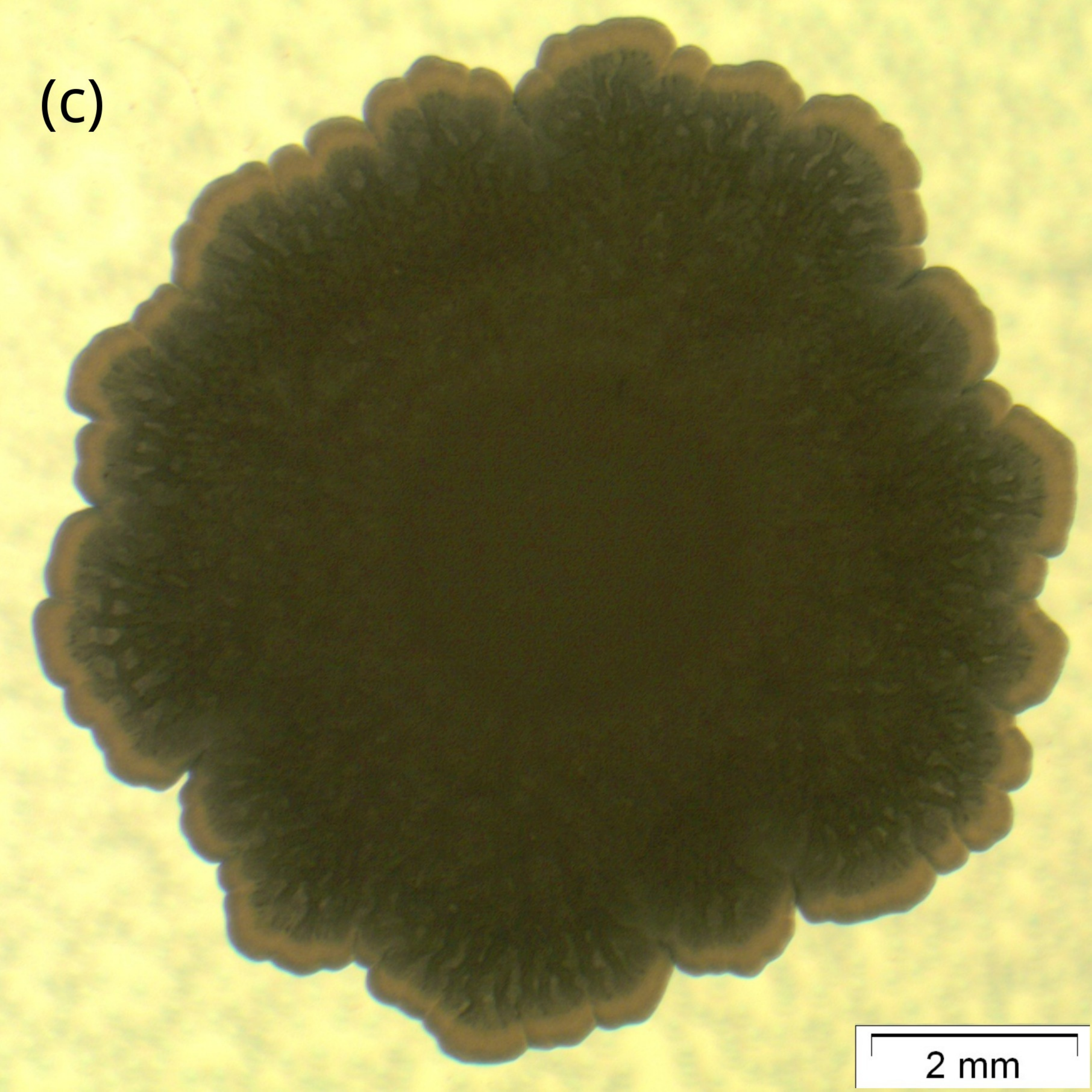}
  \includegraphics[width=0.25\textwidth]{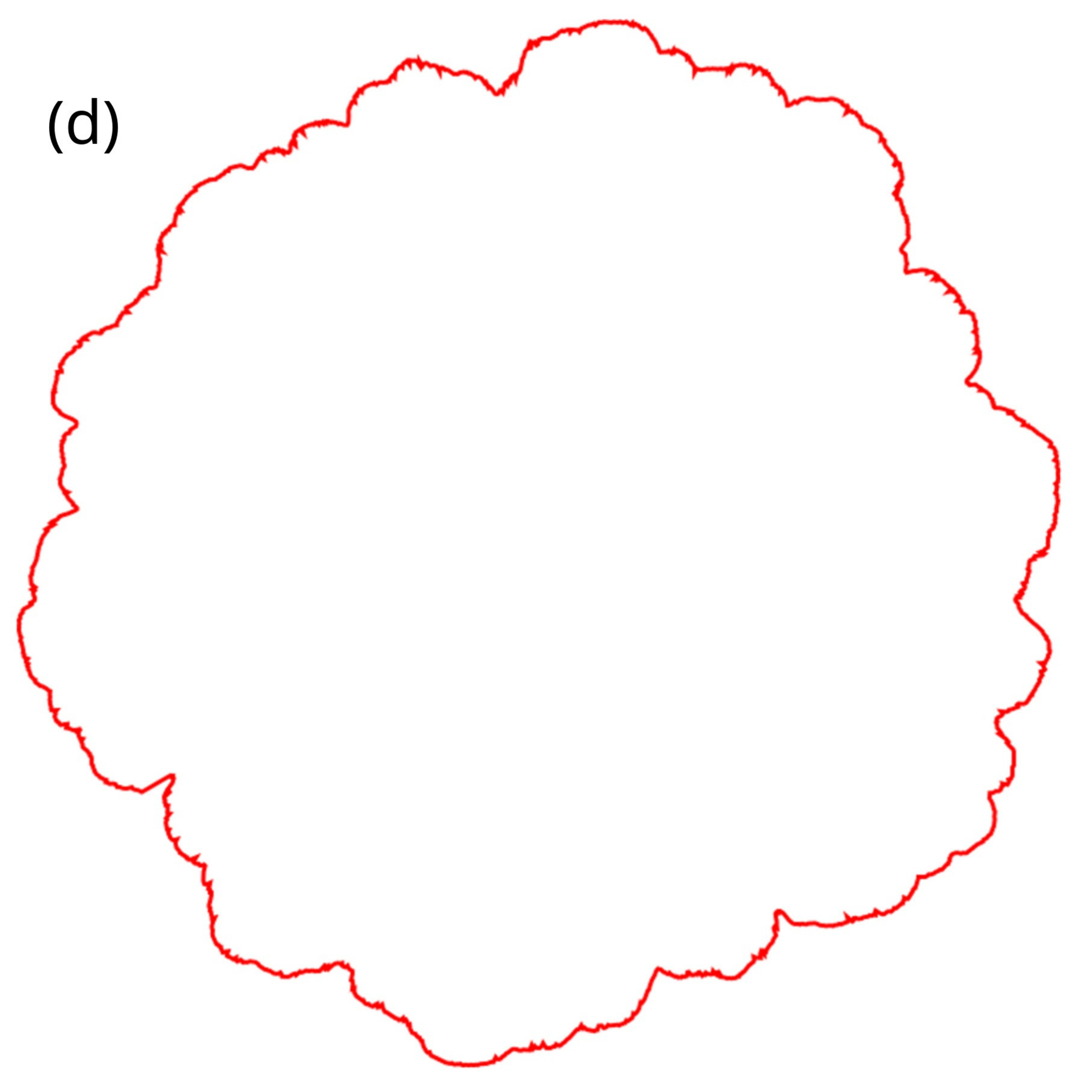}}
\\
\mbox{
  \includegraphics[width=0.25\textwidth]{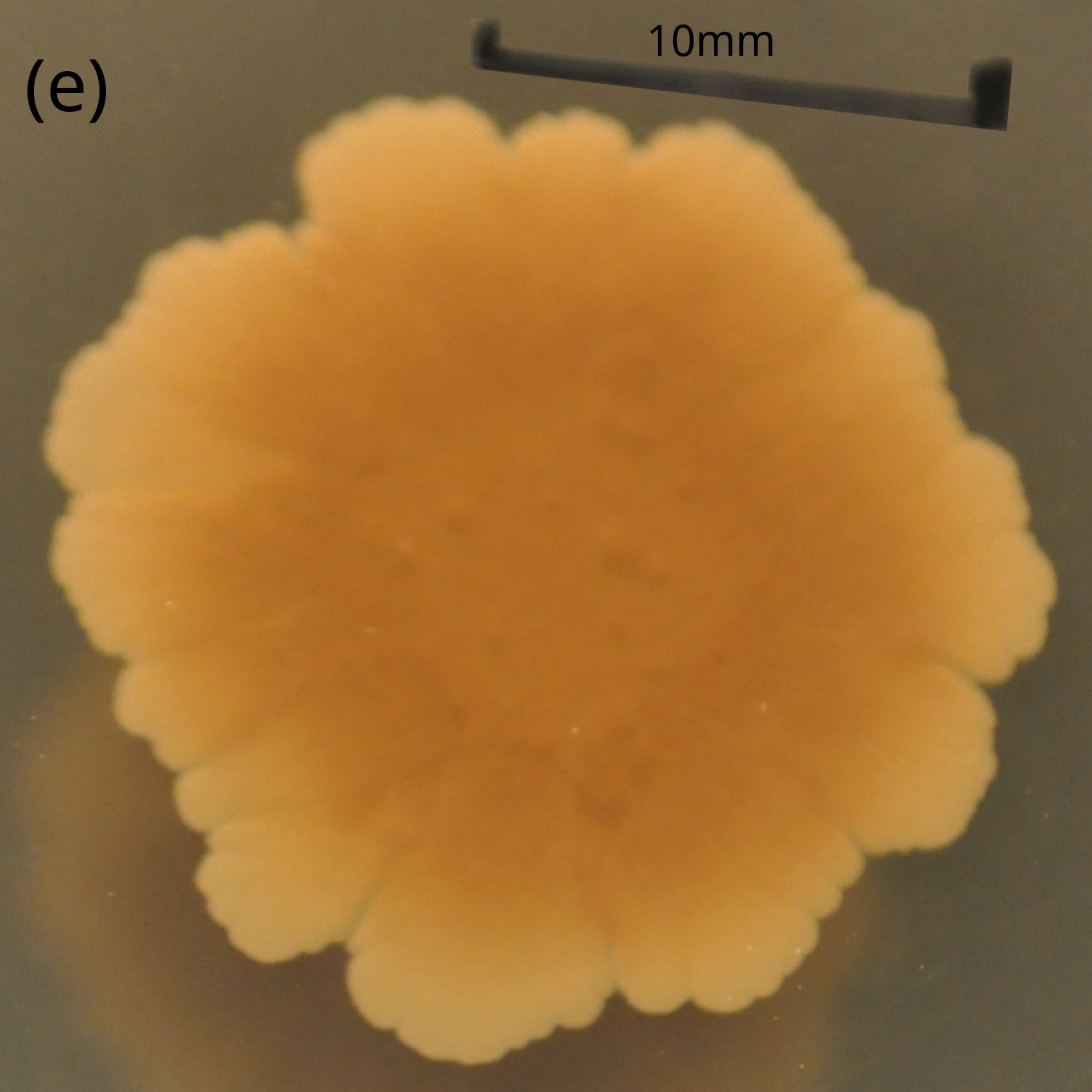}
  \includegraphics[width=0.25\textwidth]{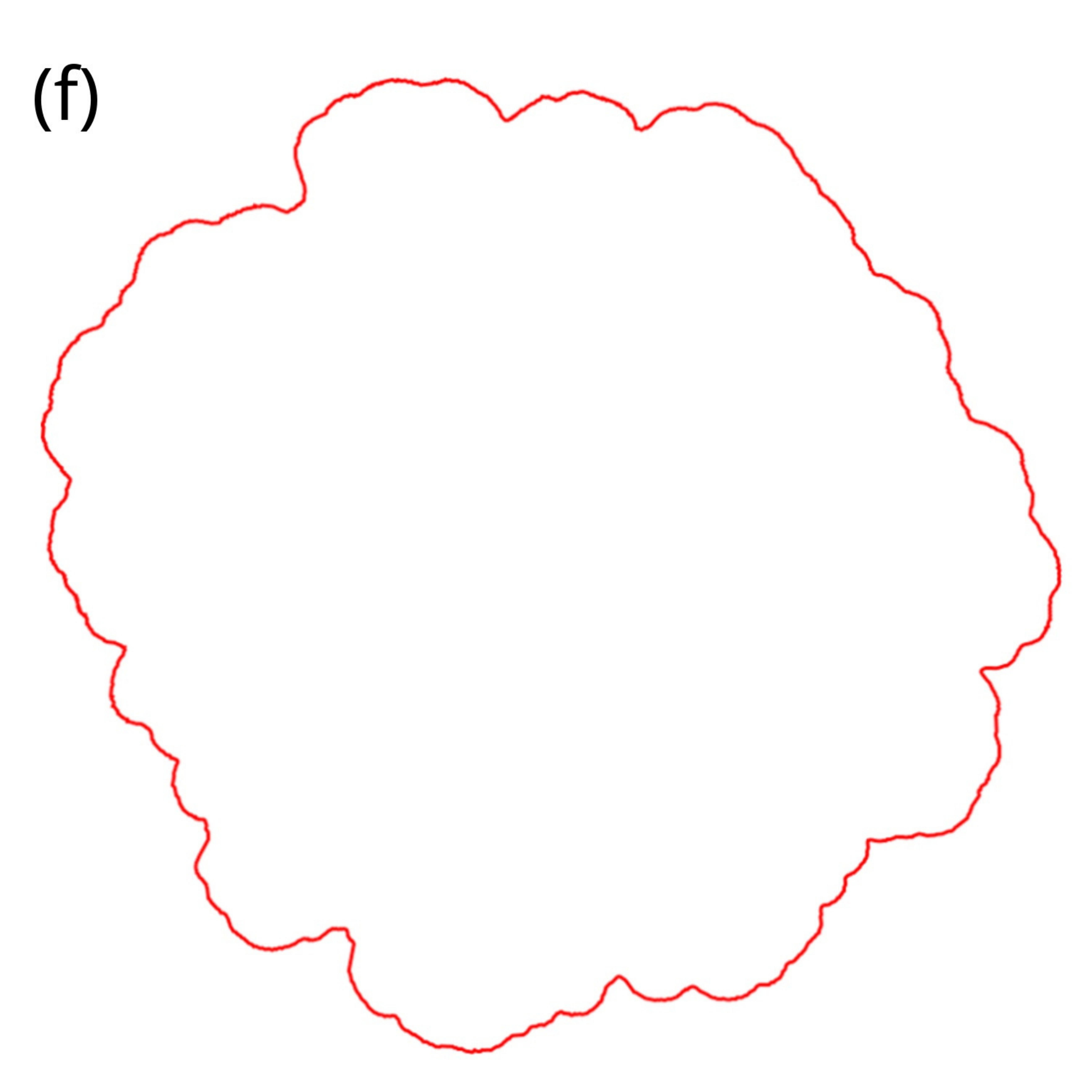}}

\caption{Experimental photographs of the bacterial
  colonies (left column, (a, c, e)) and profiles extracted using the procedure
  described in the text (right column, (b, d, f)). All these examples corresponds
  to {\em B.\ subtilis} with $C_n=20$ g/l. The growth times are: (a, b) 19 h, (c, d) 168 h, and (e, f) 792 h, top to bottom.}
\label{fig:profiles}
\end{figure}

Pictures were taken at different incubation times using a digital
camera (Olympus SC30, Japan; 3.3 Mp) coupled to a stereo microscope
(Olympus SZX10), or a digital camera (Nikon D5000, Japan; 12.3 Mp) for
large enough colonies. These photographs were treated to extract the
position of the colony front at each growth time, see
Fig.\ \ref{fig:profiles}.

\subsection{Extraction of front profiles}

We next consider the protocol that we have
followed in order to extract the position of the fronts of the
bacterial colonies from the photographs. The analysis was
semi-automatic. An algorithm was developed, which works in the
majority of the cases without supervision. The images were digitized
and subject to a constrast filter in order to highlight the
interface. The resulting image can be regarded as a matrix with entries equal to $1$ inside 
the colony and equal to $0$ outside the colony. Then, a discretized
continuous curve was obtained as follows. First, the geometric center
of the colony bulk was estimated. Then we obtained the intensity curve
along rays emanating from that point for different angles,
$I_\theta(r)$. For each angle $\theta$, we obtained the distance
$r(\theta)$ from the center, such that a certain threshold value of
the total intensity was found below it. Mathematically,
\begin{equation}
\int_0^{r(\theta)} dr \, I_\theta(r) = \mu \int_0^\infty dr \, I_\theta(r) ,
\label{eq:intensity_threshold}
\end{equation}
where $\mu$ is the threshold parameter. In our present case, $\mu=0.99$ was
employed, i.e., the radius $r(\theta)$ is defined as the first
percentile of the intensity distribution. As an illustration,
Fig.\ \ref{fig:profiles} shows a set of experimental photographs and
their corresponding profiles. Note the compact form of the bacterial
colony, delimited by a well-defined front that fluctuates around an
average circular shape.


\section{Effective model}
\label{sec:model}

The evolution of the colony front can be rationalized through a
kinetic continuum model for the dynamics of the front
position. In this model the detailed dynamics of relevant
physical fields (e.g., bacterial and nutrient densities) other than
the position, $\vec{r}(t)$, of the front itself, is neglected. The model 
is tailored so as to capture purely the form and the dynamics of the front, in a similar
way to many other instances of diffusion-limited growth, like thin solid films
\cite{Bales_90,Ojeda_00,Castro_12} or combustion fronts
\cite{Frankel_95,Blinnikov_96}, in which this type of approach has proven useful. 
Specifically, we consider
\begin{equation}
  \pl_t \vec r =
  \( A_0 + A_1 K(\vec r) + A_a \Theta_a(\vec r) + A_n \eta \) \vec n ,
\label{eq:bacterial_growth}
\end{equation}
%
%
where $\vec r$ is an interface point, $\vec n$ is the local exterior
normal, $K(\vec r)$ denotes the curvature of the interface at that
point, $\Theta_a(\vec r)$ is the local aperture angle and $\eta$ is a
zero-average Gaussian uncorrelated space-time noise. Furthermore,
$A_0$, $A_1$, $A_a$ and $A_n$ are parameters which quantify,
respectively, the relative strengths of the average growth velocity of
a planar front, surface tension, the dependence on the aperture angle,
and fluctuations. Equation \eqref{eq:bacterial_growth} is similar to
continuum models earlier put forward in the context of growth of thin
solid films limited by diffusive transport, see
e.g.\ \cite{Meakin_98}. Note that, in contrast with many works in that
field, Eq.\ \eqref{eq:bacterial_growth} applies to interfaces with an
arbitrary geometry, in particular with an average circular shape, and
is not affected by small-slope, nor no-overhang approximations. In
this sense, the model can be considered a stochastic generalization of
a previous equation put forward in the context of combustion fronts
\cite{Frankel_95,Blinnikov_96}, for which transport also takes place
by diffusion.

\begin{figure}
\begin{center}
\includegraphics[width=5cm]{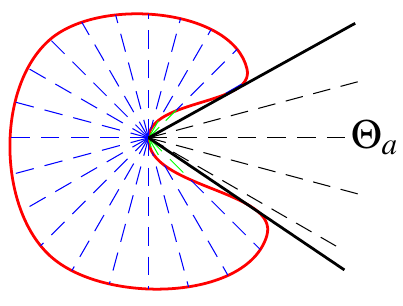}
\end{center}
\caption{Given an interface (shown by the curved red
  solid line), and a point $\vec{r}$ on it, let us consider all the rays
  emanating from this point (dashed straight lines), and find out the fraction of rays that do not intersect the interface again (those delimited on the right side of the figure by the straight black solid lines). Such a fraction provides the local aperture angle, $\Theta_a(\vec{r})$.}
\label{fig:shadowing}
\end{figure}

\begin{figure}[t!]
\centering
\includegraphics[width=0.45\textwidth]{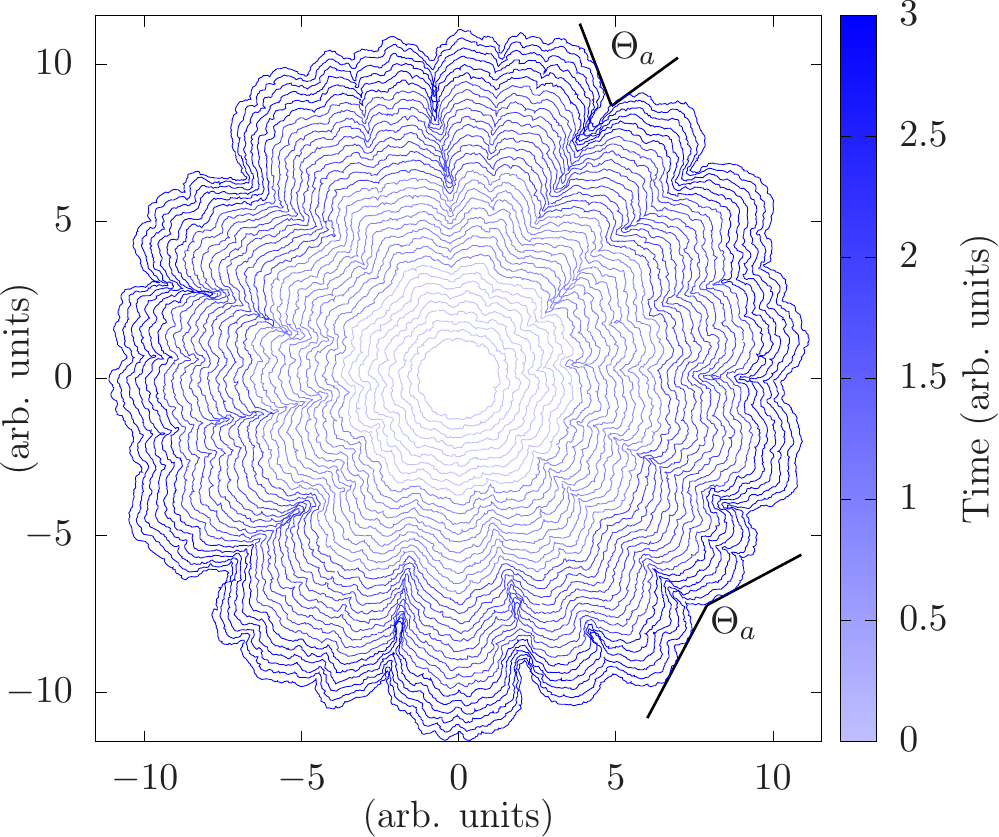}
\caption{Interfaces from
  numerical simulations of Eq.\ \eqref{eq:bacterial_growth} for
  $A_0=0$, $A_1=0.1$, $A_a=1$, $A_n=0.1$ and a circular initial
  condition. Sketches further illustrate the meaning of the local aperture
  angle $\Theta_a(\vec r)$. The growth time for each profile can be read
  from the color bar on the right. Space and time units are
  arbitrary.}
\label{fig:foto}
\end{figure}

In our model, we assume that growth resources increase locally with the angle under which a given point $\vec{r}$ at the interface sees the exterior world, which we describe as the {\em
aperture angle}, $\Theta_a(\vec{r})$, wich is illustrated by the sketch on Fig.\ \ref{fig:shadowing} and further in Fig.\ \ref{fig:foto}. Intuitively, points inside cavities get less nutrient than those at local protuberances. As frequently done in the context of diffusion-limited growth, one may make an analogy \cite{Meakin_98} to an ensemble of grass leaves which are striving to collect sunlight: taller leaves cast shadows on shorter ones, hindering growth of the latter. With this metaphor in mind, we consider this term to implement a {\em shadowing effect}, as frequently done in the context of DL growth \cite{Meakin_98}.
Mathematically, the computation of the aperture angle is performed as
follows. Let $\Gamma$ be the interface, with $\vec r_0$ and $\vec r$ being
points on it. Let $A(\vec r,\vec r_0)$ be the angle under which
$\vec r$ is seen from $\vec r_0$. Then, the aperture angle from point
$\vec r_0$ is given by
\begin{equation}
\Theta_a(\vec r_0)=2\pi -
\left|\mbox{Range}_{\vec r\in \Gamma}\left( A(\vec r,\vec r_0) \right)\right|.
\label{eq:material_angle}
\end{equation}
i.e. the measure of the range of function $A(\vec r,\vec r_0)$ when
$\vec r$ takes values on $\Gamma$.


Equation \eqref{eq:bacterial_growth} implements the basic mechanisms influencing
growth dynamics: on average, the front tends to minimize its length and
grows along the local normal direction, faster at those locations $\vec{r}$ which are more exposed [larger aperture angle $\Theta_a(\vec r)$] to the external diffusive fluxes; moreover, the front position experiences stochastic fluctuations related with microscopic events (nutrient transport and consumption, as well as cell division and relocation). The choice of these mechanisms is supported by more detailed continuum models of bacterial colonies \cite{Dockery_02,Giverso_15,Giverso_16} which find the front to be unconditionally unstable to perturbations. In particular, no quenched disorder is assumed.


In order to simulate Eq.\ \eqref{eq:bacterial_growth}, we have proceeded along the lines of \cite{Rodriguez-Laguna_11,Santalla_14}: the interface is discretized
in an adaptive way, adding and removing points dynamically in order to
keep a constant spatial resolution. The normal vector and the local
curvature are computed using concepts from discrete geometry. An
important element of the simulation is that the interface is always a
simple curve, although it can have {\em overhangs}: self-intersections
are removed.

The evaluation of the aperture angle is the most costly part of the
calculation to simulate Eq.\ \eqref{eq:bacterial_growth}, since it is
a global measurement. We have devised the following algorithm in order
to compute it. Given a point $P$ and a segment
$P_1P_2$, we define the minimal angle-interval as the counterclockwise ordered
pair $\alpha(P,P_1P_2)\equiv (\alpha_0,\alpha_1)$ of angles, with
respect to the horizontal, under which the segment is viewed from the
point. If a segment is extended to a chain $P_1\cdots P_n$, we just
compute the union of all angle-intervals. The aperture angle is the complementary of the measure of the final angle-interval.


In order to assess the type of morphological instability implied by
the aperture term in Eq.\ \eqref{eq:bacterial_growth}, we have
simulated it numerically by setting to zero all other terms in the
equation. We have performed a linear stability analysis of the ensuing
model by studying the rate of growth or decay in time for
sinusoid-like perturbations of an overall circular shape (not
shown). We have verified the expected unstable behavior: the amplitude
of a small perturbation grows with a velocity proportional to the
wave-number $k$ of the perturbation. In the case of a band geometry
with periodic boundary conditions, this means that, according to
Eq.\ \eqref{eq:bacterial_growth},
\begin{equation}\label{eq:h_k}
\partial_t h_k(t) \simeq |k| \, h_k(t) + \cdots ,
\end{equation}
where $h_k(t)$ is the amplitude of a small sinusoidal perturbation of
a flat profile with wave-vector $k$. This is indeed the well-known
behavior of the aperture-angle term, as elucidated in other
diffusion-limited systems
\cite{Meakin_98,Frankel_95,Blinnikov_96}. The growth law 
Eq.\ \eqref{eq:h_k} corresponds specifically to the destabilizing component 
of the classic Mullins-Sekerka instability, paradigmatic of diffusion-limited growth 
\cite{Vicsek_92,Meakin_98}.


Figure \ref{fig:foto} shows the time evolution of an initially
circular interface described by Eq.\ \eqref{eq:bacterial_growth}, as obtained from numerical
simulations for a representative choice of parameters. Once the
interface perimeter grows large enough, the shadowing instability indeed sets
in, reflecting the preferential growth at front protrusions, as
compared with front troughs. In strong similarity with the
experimental profile on the Fig.\ \ref{fig:profiles}, the colony
remains a compact aggregate for all $t$, with a front that fluctuates
around an average circular shape.




\section{Experimental Results}
\label{sec:analysis}

In this section we report our experimental results for BS and EC conlonies. Along with the various experimental properties studied, we additionally consider numerical simulations of Eq.\ \eqref{eq:bacterial_growth} as aids to interpret the experimental results.

\subsection{Time evolution: Radius and global roughness}

We first consider quantitatively the time evolution of our experimental BS and EC colonies through the average radius and global roughness of the colony fronts: After front extraction as described in Sec.\ \ref{sec:experiments}, each profile is a set of $N$ points on the plane, $\{x_i,y_i\}_{i=1}^N$. This set is employed to obtain the radius, $R$, of the best fitting circle, using a minimization procedure to find the corresponding center $(x_{\rm C},y_{\rm C})$. The deviations from the fitting circle provide the {\em global roughness} or surface width,
\begin{equation}
  W \equiv
  { \left< \frac{1}{N}
    \sum_{i=1}^{N}
    \left(\sqrt{(x_i-x_{\rm C})^2+(y_i-y_{\rm C})^2} - R\right)^2
    \right> }^{1/2},
\label{eq:def_W}
\end{equation}
where brackets denote averages over experimental realizations. 
\begin{figure}[t!] 
  \includegraphics[width=.5\textwidth]{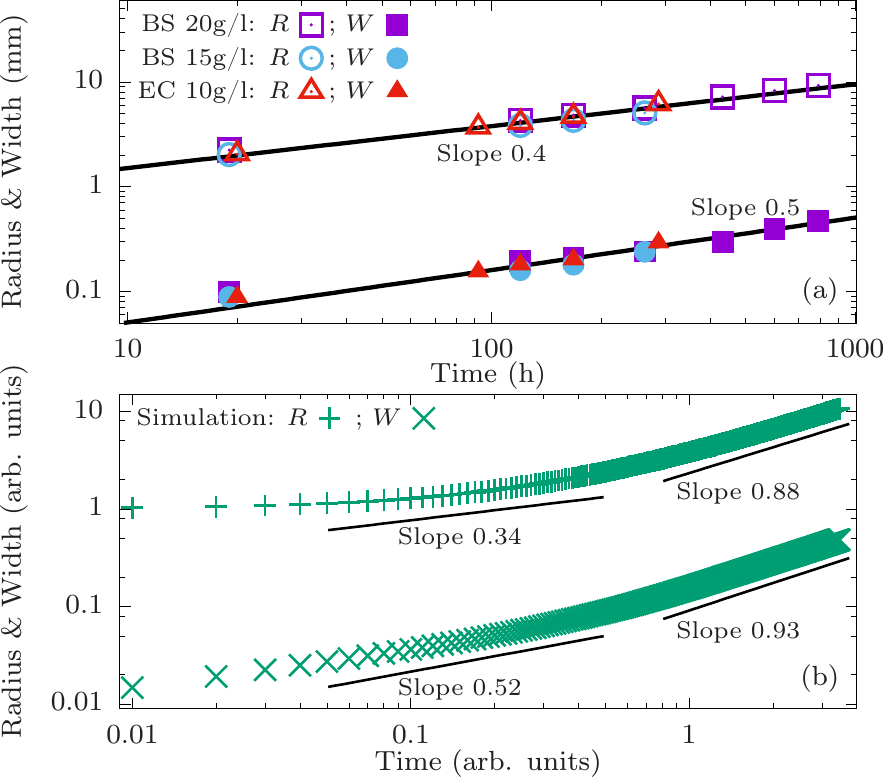}
  \caption{\label{fig:wrt} (a) Experimental radius $R$ (open
    symbols) and roughness $W$ (solid symbols) vs growth time. Purple
    and blue (red) symbols are for BS (EC), with $C_n$ as in the
    legend.
    Lines are fits to power laws, $R \sim t^n$ and $W \sim t^{\beta}$,
    with $n\approx 0.4$ and $\beta\approx 0.5$. (b) $+$
    ($\times$) symbols are data for $R$ ($W$) from numerical
    simulations of Eq.\ \eqref{eq:bacterial_growth} for parameters as
    in Fig.\ \ref{fig:foto}, averaged over 500 noise
    realizations. The lines represent power-laws $R\sim
  t^{n}$ and $W\sim t^{\beta}$ with different values of $n$ and $\beta$ for short and long times, as indicated. Units are arbitrary.}
\end{figure}
Both the radius and the global roughness of the experimental colony fronts depend on growth time. Results for $R(t)$ and $W(t)$ are provided in the top panel of Fig.\ \ref{fig:wrt}. Data can be fit by power laws in both cases, $R(t) \sim t^n$ and $W(t) \sim t^{\beta}$, with $n \simeq 0.38$--$0.43$ and $\beta \simeq 0.47$--$0.52$ values which are similar
for different nutrient concentration values and bacterial
species. Usually, for experimental circular interfaces that display
Eden/KPZ fluctuations \cite{Takeuchi_10,Yunker_13} ---conspicuously
including (Vero) cell aggregates \cite{Huergo_11}---, the average
front velocity is constant, hence the average front position increases
linearly with time. At variance with this, the radial growth rate we
measure is sublinear, i.e., $n < 1$. On the other hand, $W$ follows
power-law behavior with time as in standard kinetic roughening
systems. Taking into account that uncorrelated surface growth (so-called random deposition, RD) is characterized by $\beta_{\rm RD}=0.5$ \cite{Barabasi_95}, our relatively large experimental $\beta$ values are suggestive of uncorrelated, or possibly unstable growth
wherein front fluctuations are amplified and grow even faster than in mere RD
\cite{Vicsek_92,Barabasi_95,Meakin_98}. As noted in
\cite{Bonachela_11}, to date no other experimental work on bacterial
colony growth provides information on the time evolution of $R(t)$ or
$W(t)$ under our working conditions, in spite of the fact that
universality classes are defined by two independent exponents
\cite{Vicsek_92,Barabasi_95,Meakin_98}, one of them related with
time-dependent behavior. 

For the sake of comparison, the bottom panel of Fig.\ \ref{fig:wrt} shows the average radius and global roughness obtained from numerical simulations of our model, Eq.\ \eqref{eq:bacterial_growth}. Apparently in contrast with the experiment, for each magnitude two different regimes can be distinguished, one for short times and a different one for long times, within which the power laws are characterized by different exponent values. Note that the numerical values of the exponents which are closest to those of the experiments correspond to the model short-time regime. Actually, taking e.g.\ BS colonies with $C_n = 20$ g/l as a representative case, we can make a more detailed comparison between the experimental behavior of $W(t)$ and $R(t)$ with that predicted by Eq.\ \eqref{eq:bacterial_growth}.

\subsubsection{Simulations in physical units}
\label{sec:phys_units}

The experimental data for the evolution of the global roughness agree closely with the early time behavior of the simulations; these were performed for several sets of parameter values, with very similar results. The specific choice given in Fig.\ \ref{fig:foto} (namely, $A_0=0$, $A_1=0.1$, $A_a=1$ and $A_n=0.1$) turned out to be the most relevant one to our experimental system. Of course, the units for these constants are arbitrary in principle. However, we can convert them into physical units through detailed comparison with the experimental data, as follows.

\begin{figure}[t!] 
  \includegraphics[width=.5\textwidth]{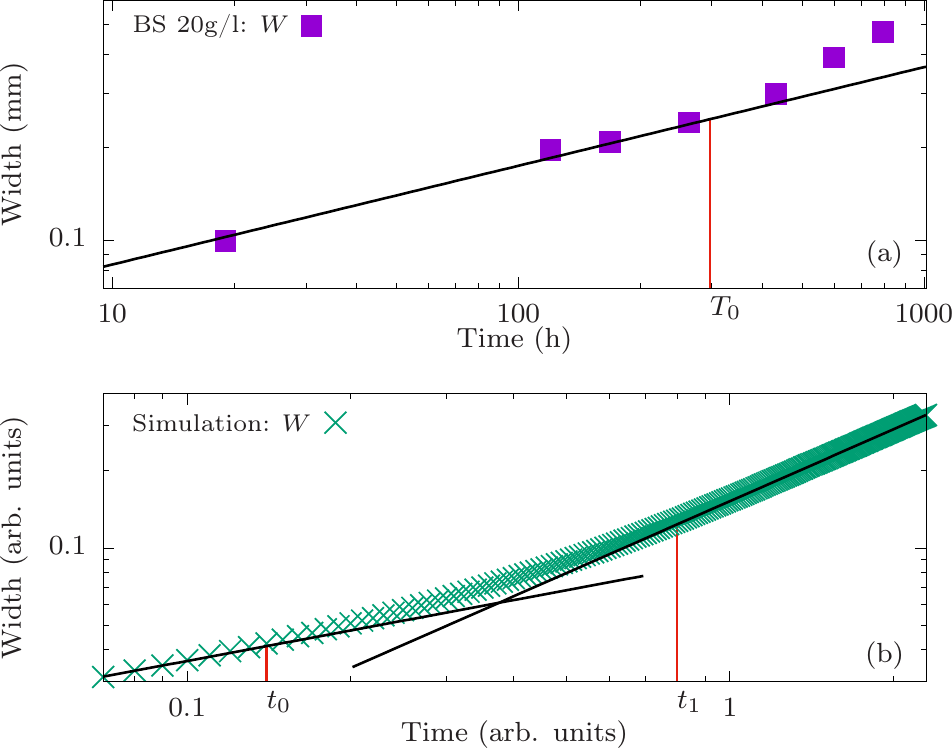}
  \caption{\label{fig:numbers} (a) Magnified view of the global roughness of a BS bacterial colony studied in Fig.\ \ref{fig:wrt}. The time marked as $T_0$ indicates a change in the power-law behavior. (b) Magnified view of the roughness from numerical simulations of Eq.\ \eqref{eq:bacterial_growth} using parameters as in Fig.\ \ref{fig:wrt}. Time $t_0$ corresponds to the initial change in scaling behavior. Time $t_1$, signalling the beginning of asymptotic, long-time behavior, is also indicated.}

\end{figure}
\begin{figure}[t!] 
  \includegraphics[width=.48\textwidth]{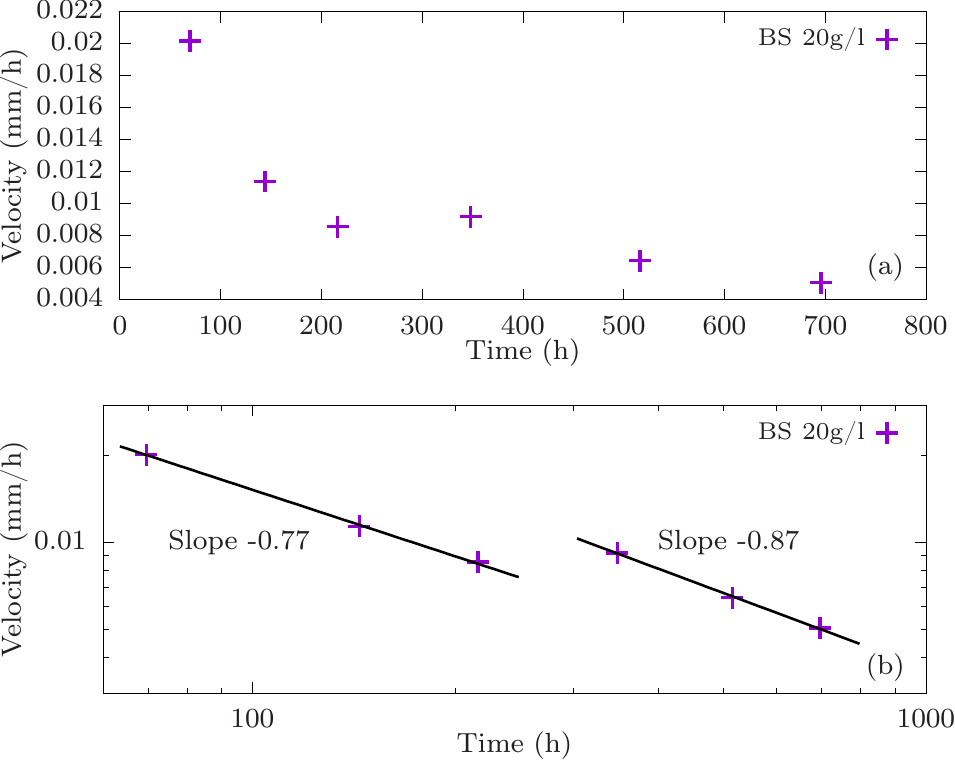}
  \caption{\label{fig:speed} Average front speed as a function of time for BS experiments using  $C_n=20$ g/l in linear (a) and doubly-logarithmic (b) displays. The data group themselves into two scaling regimes, approximately separated at $T_0=297$ h.}
\end{figure}

In Fig.\ \ref{fig:numbers}(a) we show the roughness of the interface, $W(t)$, for the same {\em B.\ subtilis} experiments with $C_n=20$ g/l considered in Fig.\ \ref{fig:wrt}, but in a magnified view. A certain time $T_0=297$ hour (h) can be identified which marks a change in the power-law behavior of the data, at which the global roughness is $W_0=0.25$ mm. The experiment ends at time $T_e=801$ h, when $W_e=0.47$ mm. Thus, we have $W_e/W_0=1.9$ and $T_e/T_0=2.7$. The physical occurrence of $T_0$ can be confirmed by in other measurable quantitites, such as the average front velocity, see Fig. \ref{fig:speed}. 
The front speed is estimated by comparing consecutive measurements of the radius and
using a finite-differences approach. The two panels show the same data, the only difference between them being that the bottom one is shown in logarithmic scale. We can see how the data divide into two sequences of points with slightly different scaling behavior, with the
division approximately corresponding to $T_0=297$ h.

\begin{figure}[t!] 
  \includegraphics[width=.45\textwidth]{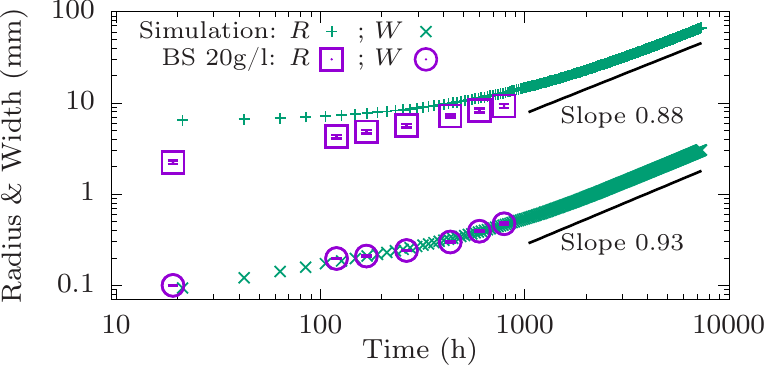}
  \caption{\label{fig:wrt2} Evolution of the radius and global roughness predicted by Eq.\ \eqref{eq:bacterial_growth} with physical parameters $A_0=0$ mm/h, $A_1=0.067$ mm${}^2$/h, $A_a=6.1\cdot 10^{-3}$ mm/h, $A_n=0.086$ mm${}^{3/2}$/h${}^{1/2}$. Squares (circles) are experimental radius (roughness) for BS with $C_n=20$ g/l; error bars are of the same size as symbols or smaller, see appendix A.}
\end{figure}

Coming back to the simulations of Eq.\ \eqref{eq:bacterial_growth}, Fig.\ \ref{fig:numbers}(b) indicates a change in the scaling behavior of the global roughness at time $t_0=0.14$ [T], with a roughness of $w_0=0.044$ [L], where [L] and [T] are length and time units, respectively. Thus, the end of the experiment should correspond to a roughness
$w_e=0.044$ [L] $\times 1.9=0.084$ [L], which takes place at $t_e\sim
0.44$ [T]. We make this time correspond to $T_e=801$ h. Thus, the
numerical conversion from arbitrary time units to hours is $801$ h$/0.44$ [T] $\approx 1800$ h/[T]. The same reasoning can be performed with the length unit and we obtain a conversion factor of $0.47$ mm$/0.044$ [L] $\approx 11$ mm/[L]. Alternative procedures can be designed to obtain the conversion factors, and they all provide similar results. At any rate, using the indicated conversion factors we can estimate the physical values of the equation parameters in physical units, namely, $A_0=0$ mm/h, $A_1=0.067$ mm${}^2$/h, $A_a=6.1\cdot 10^{-3}$ mm/h, $A_n=0.086$ mm${}^{3/2}$/h${}^{1/2}$. Experimental data are compared with simulations for this parameter choice in Fig.\ \ref{fig:wrt2}.
With respect to $W(t)$, agreement is reached for essentially the full duration of the experiments. For times longer than approximately $800$ hours (which remain beyond our experimental setup), Eq.\ \eqref{eq:bacterial_growth} predicts almost linear increase
with time for $W(t)$ and $R(t)$. The agreement between experimental and
simulation data is slightly worse in the case of $R(t)$, for which the
initial condition plays a stronger role in the continuum model. 
Nevertheless, agreement also becomes quantitative for $t > 100$
hours. Note, the time $t_1$ required for the onset of long-time, asymptotic behavior in the experiments can be assessed from the numerical simulation of Eq.\ \eqref{eq:bacterial_growth}, see the bottom panel of Fig.\ \ref{fig:numbers}. We estimate $t_1=0.8$, which approximately corresponds to $1440$ hours. Overall, Fig.\ \ref{fig:wrt2} suggests that the scaling behavior reached in the experiments is preasymptotic, clear-cut asymptotics occurring for $t > t_1$, approximately twice our longest experimental growth time.

\subsection{Geometrical properties: Local roughness and radial correlations}

\begin{figure*}[t!]
\includegraphics[width=\textwidth]{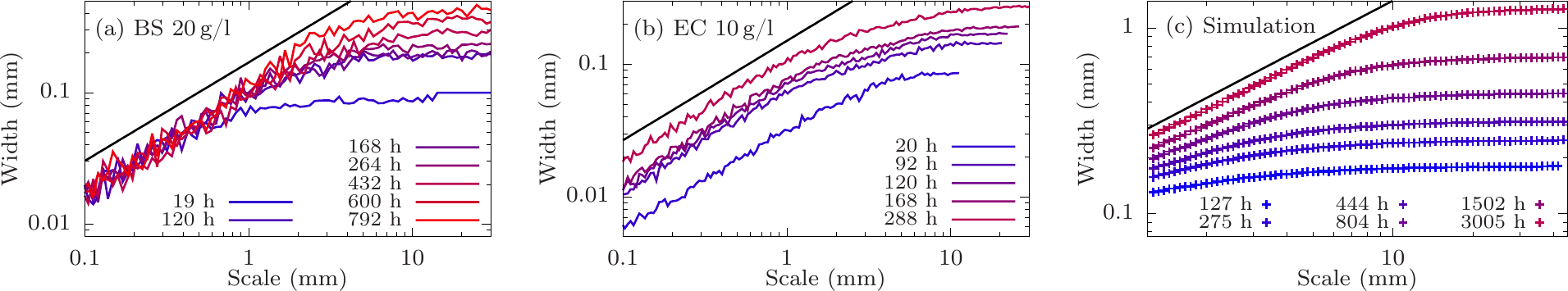}
\caption{\label{fig:morphology} Experimental length-scale dependence of
  the local roughness of BS (a) and EC (b) colony fronts for different times and $C_n$ as indicated in the legends. (c) Same observable for numerical simulations of Eq.\ \eqref{eq:bacterial_growth} as in Fig.\ \ref{fig:foto}. All straight lines represent $w(\ell) \sim
  \ell^{0.75}$.}
\end{figure*}

\begin{figure*}[t!]
\includegraphics[width=\textwidth]{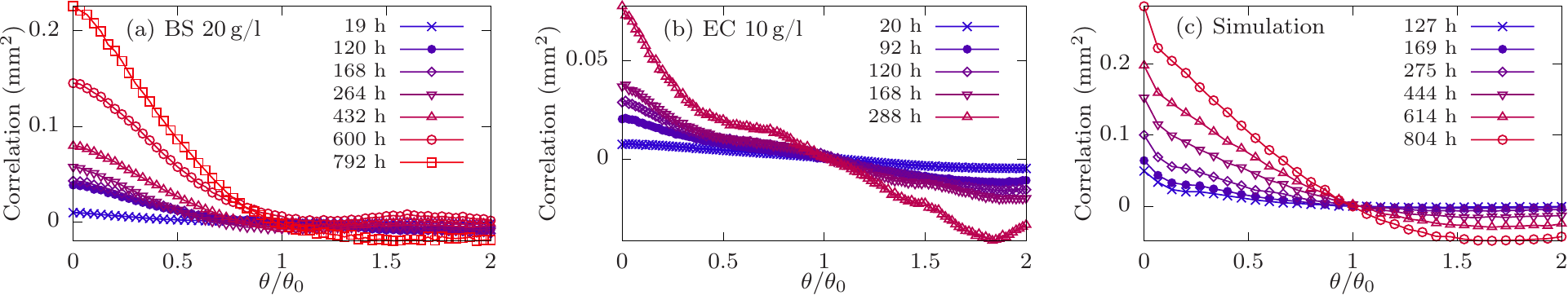}
\caption{\label{fig:corr} Left to right: Autocorrelation function of experimental radial fluctuations vs angular distance rescaled by $\theta_0$, for BS (a), EC (b), and simulations (c) of Eq.\ \eqref{eq:bacterial_growth} with parameters as in Fig.\ \ref{fig:foto}, where $\theta_0=30^\circ, 60^\circ$, and $15^\circ$, respectively. Times and $C_n$ are as given in the legends.}
\end{figure*}

\begin{figure}
\includegraphics[width=.45\textwidth]{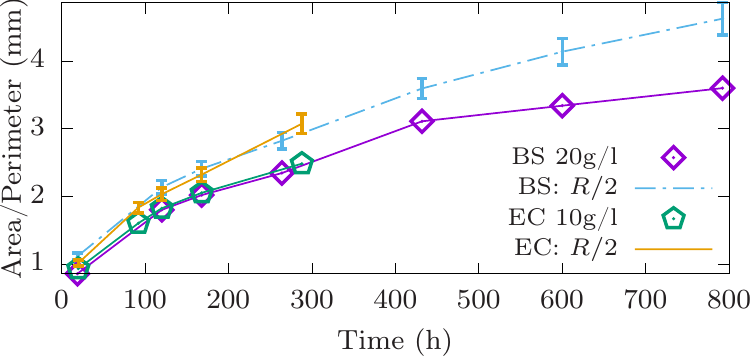}
\caption{\label{fig:arearadio} Experimental area/perimeter ratio vs time. Symbols are direct measurements for conditions described in the legend; lines show $R(t)/2$ as obtained from Fig.\ \ref{fig:wrt} for the corresponding sets of data.}
\end{figure}

Further non-trivial properties of the experimental colonies involve the spaial dependence of front fluctuations. We can characterize them quantitatively by considering the so-called local roughness, $w(\ell)$, which evaluates interface deviations from an average position, within observation windows of size $\ell$ \cite{Vicsek_92,Barabasi_95,Meakin_98}. We proceed as is customary for systems with an overall circular symmetry \cite{Meakin_98,Santalla_14}: Namely,
each point on the front is converted to polar coordinates emanating
from the geometric center, $(x_i,y_i)\to (\theta_i,r_i)$, whereby
$\theta_i$ ($r_i$) is considered a new independent (dependent)
variable. Given an initial point $\vec r_0$ and a length scale $\ell$,
we consider the set of points within a circle centered at $\vec r_0$
with radius $\ell$. Then, we make a fit to the straight line which
minimizes the deviations. The mean-square distance of the front
positions to that fitting line provides the {\em local} roughness
$w(\ell)$.
Results for our experimental BS and EC colonies are displayed in Fig.\ \ref{fig:morphology}.
An approximate power-law dependence, $w(\ell)\sim \ell^{\alpha}$, holds at intermediate scales above 100 $\mu$m, and up to $3$ mm for the most favorable cases, with $\alpha\simeq 0.75$.
For the sake of comparison, we recall that a one-dimensional interface provided by the world-line of an uncorrelated random walk features $\alpha_{\rm RW}=1/2$ \cite{Vicsek_92,Barabasi_95,Meakin_98}. Our experimental value for $\alpha$ is in the same range as those found earlier for
similar bacterial colony experiments \cite{Vicsek_90,Wakita_97,Bonachela_11} and is also similar to values measured in other DL systems, like 1D ECD \cite{Pastor_96,Huo_01} or 2D thin films grown by CVD under low sticking conditions \cite{Ojeda_00,Zhao_00}. 
In these contexts, such large $\alpha$ are known not to correspond to any well-defined universality class of kinetic roughening \cite{Castro_00,Nicoli_09,Castro_12}, but to merely reflect the large surface slopes that ensue, due to diffusive instabilities. Such instabilities are actually well-known to correlate with front branching \cite{Vicsek_92,Barabasi_95,Meakin_98}, which in our experiments can be assessed through the behavior of the autocorrelation of the radial interface fluctuations as a function of the angular distance, 
\begin{equation}
C(\Delta\theta,t) = \langle [r(\theta,t)-R(t)][r(\theta+\Delta\theta,t)-R(t)]\rangle .
\label{eq:C}
\end{equation}
As seen in Fig.\ \ref{fig:corr}, and in spite of the compactness of the colonies, $C(\Delta\theta,t)$ vanishes approximately at the same angular distance for different times, indicating fronts that develop well-defined branches. Moreover, the importance of such branching increases monotonically along the experimental time evolution. Such a behavior is analogous to the result of detailed continuum models of bacterial colony growth put forward
in \cite{Giverso_15,Giverso_16}, which predict unconditional instability of the colony front to perturbations for a variety of relaxation mechanisms that include both, chemotactic and volumetric expansions. In application of the analysis in \cite{Giverso_15,Giverso_16} to our data, Fig.\ \ref{fig:arearadio} shows the time evolution of the area/perimeter ratio for the experimental colonies, compared to the $R(t)/2$ value that would correspond to a perfectly circular front in each case; clearly the actual perimeter grows too fast with time relative to the enclosed area, as compared with expectations for an ideally circular front. Such a behavior is inconsistent in particular with the occurrence of Eden behavior at long times \cite{Vicsek_92,Barabasi_95,Meakin_98}.

%

The geometrical properties of the front observed in the experiments are very similarly found also in the simulations of Eq.\ \eqref{eq:bacterial_growth}. Figure \ref{fig:morphology} shows the dependence of the simulated local roughness with length scale for different times, readily compared with the experimental data in the same figure. Indeed, for small scales the model yields $w(\ell)\sim \ell^{\alpha}$, with $\alpha$ increasing with time up to 0.75, very close to the experimental value. Comparison between the model and the experiments improves with increasing times, as typical length-scales also increase. Note, these simulation data include the long-time, asymptotic regime identified for the model in Fig.\ \ref{fig:wrt}. Finally, the behavior in simulations of the radial autocorrelation function $C(\Delta\theta,t)$ also supports our interpretation on branching at the interface: Figure \ref{fig:corr} indeed shows the selection of a precise correlation angle value $\theta_0$, analogous to the experimental morphologies.

%
%
%


\section{Discussion and Conclusions} 
\label{sec:conclusions}

Summarizing our experimental observations for both BS and EC colonies in the suppressed-motility conditions \cite{Wakita_97,Matsushita_98,Rafols_98} which are in the classically alleged Eden regime, we obtain branched interfaces with scaling exponents $\beta\simeq0.5$ and $\alpha\simeq0.75$, which unambiguously differ from Eden/KPZ behavior, characterized by non-branched interfaces, $\beta_{\rm KPZ}=1/3$, and $\alpha_{\rm KPZ}=1/2$ \cite{Vicsek_92,Barabasi_95,Meakin_98}. Our experimental data are also inconsistent with quenched noise effects which, e.g., allegedly induce $\beta=0.61, \alpha=0.68$ in agent-based simulations of bacterial colonies \cite{Bonachela_11}, or with the so-called quenched KPZ (qKPZ) equation \cite{Barabasi_95}. Unconditioned by any comparison to models, the fact that our experimental colony profiles become increasingly branched during all accessible times (Figs.\ \ref{fig:corr} and \ref{fig:arearadio}) moreover suggests that the observed scaling is preasymptotic behavior for a system whose asymptotics is not Eden, and we speculate that this could also be the case for other, classical, experiments \cite{Vicsek_90,Wakita_97} performed under conditions which are similar to ours.

Given the semi-quantitative agreement between our experiments and simulations of the effective
model, Eq.\ \eqref{eq:bacterial_growth}, we can consider the latter in order to predict what would be the actual asymptotic behavior of the former. Indeed, Eq.\ \eqref{eq:bacterial_growth} predicts a long-time behavior with $\beta=0.93$ (Fig.\ \ref{fig:wrt}) and $\alpha=0.75$ (Fig.\ \ref{fig:morphology}). Actually, a small-slope approximation of Eq.\ \eqref{eq:bacterial_growth} yields dimension-independent exponents $\alpha=\beta=1$ \cite{Nicoli_09,Nicoli_09b} ---recently measured in CVD under DL conditions \cite{Castro_12}---, which are definitely non-KPZ and are expected to characterize Eq.\ \eqref{eq:bacterial_growth} at long times. Note, for interfaces with $\alpha\gtrsim 1$, local measurements using $w(\ell)$ are known to underestimate the correct value of the roughness exponent \cite{Castro_00}, explaining our $\alpha=0.75$ value. Parameter conditions in our experiments would make such a long-time regime hardly accessible, requiring growth times at least twice the longest time that we have been able to reach, as estimated in Sec.\ \ref{sec:phys_units}. On the other hand, the preasymptotic ($t< 800$ h) behavior in Eq.\ \eqref{eq:bacterial_growth} ---during which $W(t)$ evolves as in our experiments--- is dominated by the diffusive (shadowing) instabilities that induce branching of the front and large exponent values. In such a case, and as shown for other DL systems \cite{Nicoli_09,Castro_00,Castro_12}, the exponent values are non-universal and may depend on parameter values and even on the specific space/time ranges in which power-law fits are attempted.


In conclusion, bacterial colonies where individual motility is
suppressed form compact aggregates whose front morphology can still be
dominated by diffusive instabilities. For our experimental conditions, similar to those in
\cite{Wakita_97,Matsushita_98,Rafols_98}, preasymptotic
scaling seems to occur at the accessed times, which in any case is not in the Eden/KPZ universality class. There is no need to invoke quenched disorder to account for this
discrepancy. Rather, the shadowing instability induces large front
fluctuations with non-universal scaling. This behavior is strongly
reminiscent of many other experimental systems
\cite{Pastor_96,Ojeda_00,Zhao_00,Huo_01,Yunker_13} in which transport-induced instabilities induce effective scaling. In some of these cases
\cite{Yunker_13} the observed kinetic roughening properties have also been associated with the qKPZ universality class, due to accidental similarities in the values of the scaling
exponents \cite{Nicoli_13,Yunker_13_2,Oliveira_14}. Note, attributing a set of scaling exponents to a well-defined, asymptotic, universality class like qKPZ, or to non-universal preasymptotic behavior as we are presently advocating for, are conceptually very different interpretations.

Non-KPZ exponents due to diffusive instabilities are also predicted by
agent-based simulations \cite{Farrell_13,Farrell_17} for small values
of the active layer thickness $\delta$. However, for sufficiently
large $\delta$ very compact colonies with extremely flat fronts are
found \cite{Farrell_13,Farrell_17}. While this seemingly questions the
prevalence of diffusive instabilities, continuum models
\cite{Giverso_15,Giverso_16} analytically predict such flat front
conditions to be a finite-size effect. Thus, parameter conditions
select a typical length-scale $\ell_0$ for the instabilities,
which is well defined for any value of $\delta$. As
standard in pattern formation \cite{Cross_09}, the correlation length
along the front (initially a few cell sizes across) needs to increase
up to $\ell_0$ for the instability to set in. If $\ell_0$ is very
large (in band geometry, for systems smaller than $\ell_0$), the front
may effectively be flat.  In circular geometries, for sufficiently
(perhaps, exceedingly) long times, the instability will still occur.

We should still note that additional systems exist, which are closely
related to the ones we study, and for which Eden/KPZ scaling does
occur. For instance, bacterial colonies for which individual motility
is non-negligible \cite{Wakita_97} yield a roughness exponent
compatible with the 1D KPZ value. Also, aggregates of non-cancerous
(Vero) or cancerous (HeLa) primate cells display unambiguous KPZ
\cite{Huergo_11}, and even qKPZ \cite{Huergo_14,Muzzio_16}, scaling,
as is the case with fungal growth \cite{Lopez_02}. Experimentally, KPZ
scaling also applies to fluctuating frontiers between different genetic
strains in range expansion of {\em E.\ coli} \cite{Hallatschek_07},
although deviations from Eden behavior can also occur
\cite{Kuhr_11,Reiter_14}. In general, individual cell motility seems
to play a relevant role, to the extent that instabilities associated
with nutrient transport can eventually be superseded. Indeed, the Eden
model \cite{Eden_61} will at any rate stand as the prime example for
reaction-limited growth \cite{Vicsek_92,Barabasi_95,Meakin_98}, where
nutrient transport is, effectively, infinitely fast and irrelevant to
front fluctuations.


\begin{acknowledgments}
We acknowledge fruitful conversations with G.\ Melaugh and
K.\ A.\ Takeuchi. This work has been supported by Ministerio de Economía y 
Competitividad, Agencia Estatal de Investigación, and Fondo Europeo de 
Desarrollo Regional (Spain and European Union) through grants FIS2015-66020-C2-1-P, FIS2015-69167-C2-1-P, FIS2015-73337-JIN, and BIO2016-79618-R, and by Comunidad Aut\'onoma de Madrid (Spain) Grant NANOAVANSENS S2013/MIT-3029.
\end{acknowledgments}



\appendix

\section{Some error estimates}
\label{sec:appendix}

For each bacterial species and nutrient concentration we have only one
sample available. Therefore, the error bars on the measurements of the
radius and the roughness can not be estimated via statistical error
between different samples. An alternative approach is to consider that
measurements performed on different parts of the interface constitute
a suitable statistical ensemble from which we can estimate the
magnitude of the desired fluctuations. Thus, the global roughness itself
provides an estimate for the uncertainty in the measurement of the radius.

\begin{figure}[ht!]
  \begin{center}
    \includegraphics[width=8cm]{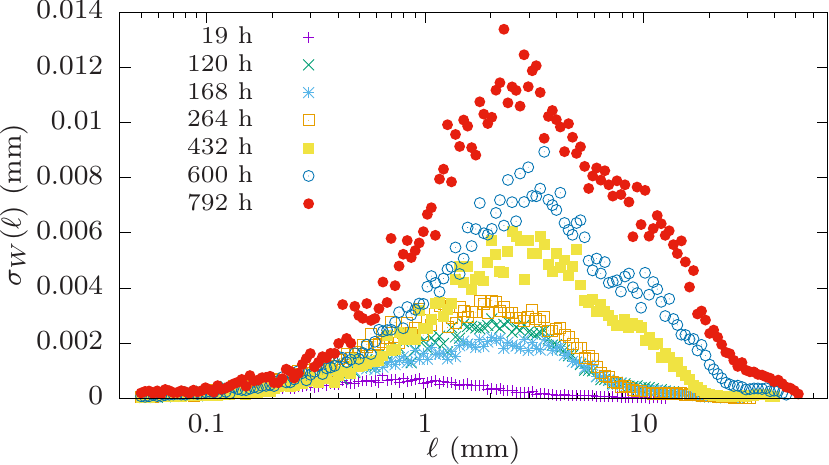}
  \end{center}
  \caption{Scale dependence of the estimate for the error in the global
    roughness obtained using Eq.\ \eqref{eq:sigmawell}, when applied to
    the profiles of our BS sample grown with $C_n=20$ g/l.
    The actual estimate is provided by the maximum of each curve.}
  \label{fig:errorbars}
\end{figure}

The estimate of the fluctuations of the roughness is more
involved. Our approach is to divide the interface into patches of
size $\ell$ and measure the estimate for each of them,
$W_i(\ell)$. Then, for each size $\ell$ we can determine the deviation
of those measures,
\begin{equation}
  \sigma^2_W(\ell)=\left\langle W^2(\ell)\right\rangle-
  \left\langle W(\ell)\right\rangle ^2.
  \label{eq:sigmawell}
\end{equation}
Naturally, this deviation will depend on the measurement scale $\ell$. We
have thus chosen the worst case to determine our estimate for the
error in the roughness, namely,
\begin{equation}
  \sigma^2_W = \max_\ell \sigma^2_W(\ell).
\end{equation}
This is how the error bars are estimated in Fig.\ \ref{fig:wrt2}. The behavior of $\sigma_W(\ell)$ for the fronts of BS colonies grown with $C_n=20$ g/l is shown as an illustration in Fig.\ \ref{fig:errorbars}.

%

\end{document}